\documentclass[twoside]{article}
\usepackage{fleqn,espcrc2}
\usepackage{graphicx}
\usepackage{here}

\newcommand{\AmS}{{\protect\the\textfont2
    A\kern-.1667em\lower.5ex\hbox{M}\kern-.125emS}}										 
\def\beq{\begin{equation}}
\def\eeq{\end{equation}}
\def\bea{\begin{eqnarray}}
\def\eea{\end{eqnarray}}
\def\bq{\begin{quote}}
\def\eq{\end{quote}}

\def\nnb{\nonumber}
\def\ga{\left(}
\def\dr{\right)}

\def\rar{\rightarrow}
\def\lrar{\Longrightarrow}
\def\llar{\Longleftarrow}																 
\def\nnb{\nonumber}
\def\la{\langle}
\def\ra{\rangle}
\def\nin{\noindent}
\def\ba{\vspace*{-0.2cm}\begin{array}}
\def\ea{\end{array}\vspace*{-0.2cm}}

\def\b{$\bullet~$}
\def\als{\alpha_s}

\def\beq{\begin{equation}}
\def\enq{\end{equation}}
\def\beqa{\begin{eqnarray}}
\def\enqa{\end{eqnarray}}

\def\G3{\lag g^3G^3\rag}

\newcommand{\rag}{\rangle}
\newcommand{\lag}{\langle}

\begin{document}
\title
{\bf{\boldmath
{\Large Duality  between QCD Perturbative  Series and Power Corrections} }}

\author{S. Narison\,\address{Laboratoire de Physique Th\'eorique et Astroparticules, CNRS-IN2P3 \&
Universit\'e de Montpellier II, Case 070, Place Eug\`ene, 34095- Montpellier Cedex 05, France}\,\thanks{{\it E-mail addresses :} snarison@yahoo.fr (S. Narison),  xxz@mppmu. mpg.de (V.I. Zakharov)}
\,\thanks{Corresponding author.},
V.I. Zakharov\address {Max-Planck-Institut f\"ur Physik, Foehringer Ring 6, 80805 Munich Germany;\\
Institute of Theoretical and Experimental Physics,
B.Cheremushkinskaya 25, Moscow, 117218, Russia.\\
}
}

\begin{abstract}
\nin
We elaborate on the relation between perturbative and power-like corrections
to short-distance sensitive QCD observables. We confront theoretical expectations with explicit
perturbative calculations existing in literature. As is expected,  the quadratic
correction is dual to a long perturbative series and one should use one
of them but not both. However, this might be true only for very long
perturbative series, with number of terms needed in most cases exceeding 
the number of terms available. What has not been foreseen, the quartic corrections
might also be dual to the perturbative series.
If confirmed, this would imply a crucial modification of the dogma.
We confront this quadratic correction 
against existing phenomenology (QCD (spectral) sum rules scales, determinations of light quark masses and of
 $\alpha_s$ from $\tau$-decay).
 We find no contradiction and (to some extent) better agreement with the data and with recent lattice calculations. 
\end{abstract}
\maketitle

\section{Introduction}
\nin
Because of the asymptotic freedom,  predictions for short-distance
processes are very simple in QCD and essentially reduce to parton model,
or to lowest order perturbation theory. This is true, however, only in
the leading order approximation. As far as corrections are concerned,
there is a double sum which includes expansion in $\alpha_s(Q^2)$
where $Q^2$ is a generic large mass parameter and powers of
$(\Lambda_{QCD}/Q)^k$. Consider for example the best studied case of current correlators
which determine QCD sum rules \cite{svz} (for a review, see e.g. \cite{SNB}).
Then, one usually assumes the following form of the correlator in the $x$-space:
\begin{eqnarray}\label{ope}
\langle 0|J(x),J(0)|0\rangle~\approx~C_I(\alpha_s(x))I~\\\nonumber
+~C_{G^2}(\alpha_s(x))G^2(0)x^4~+~...,
\end{eqnarray}
where $J(x)$ is the hadronic current, $I$ is the unit operator and $G^2$ is the dimension four operator.
The coefficient functions
$C_{I,G^2}$ are  calculable perturbatively
as infinite sums in the running coupling.

Moreover, Eq. (\ref{ope}) does not apparently contain quadratic corrections,
while such corrections are included in many cases on the phenomenological
grounds (see in particular \cite{linear,CNZ,SNe,SNtau,SNms,SNmass,ARRIOLA}).
These quadratic corrections and their
phenomenological significance will  be in fact focus of our attention.
Let us remind the reader what is understood by these corrections.

Start with the heavy quark potential at short distances.
The Cornell  version of this potential (which describes the lattice data very well) is
very  simple:
\begin{equation}\label{ya}
V_{Q\bar{Q}}(r)~\approx~-{4\over3}{\alpha_s\over  r}~+~{\sigma\cdot r}~~,
\end{equation}
where $r$ is the distance, $\alpha_s\equiv g^2/(4\pi)$ is the QCD coupling, $\sigma\approx$ 0.2 GeV$^2$ is the string tension.
The fit in Eq. (\ref{ya}) works well at all distances.
The  question is whether such a form of the potential at {\it short} distances
-- let it be only approximate -- is acceptable theoretically.
There are papers which ascertain a positive answer
to this question (see, in particular, \cite{linear,CNZ}).
The observable (heavy-quark potential in our case) is viewed as represented by a
short perturbative series (a single $const/r$ term in our case) plus a leading power correction
(quadratic correction, in our case, $\sigma \cdot r$).

The version  used in some other papers (see, in particular, \cite{brambilla}) looks as:
\begin{equation}\label{nora}
\lim_{r\to 0}V_{Q\bar{Q}}(r)~\approx~{1\over r}\sum_{n=1}^{n=4}a_n\alpha^n_s(r)~+~(const)+\tilde{\sigma}_n\cdot r~,
\end{equation}
where $n=4$ is the realistic number of perturbative terms calculated  explicitly
and  $const$ stands for an infrared  renormalon contribution
(this could be added to the version in Eq. (\ref{ya}) as well). The last term, proportional to $\tilde{\sigma}_n$
imitates the power correction.

It is quite common \cite{brambilla} to identify the parameters $\sigma$ from Eq (\ref{ya}) and
$\tilde{\sigma}$ from Eq (\ref{nora}) and compare their numerical values.
Our point is that such an identification is not justified
\,$^1$\footnotetext[1]{The conjecture was made first in a conference talk \cite{vz03}.
Present, original paper also includes material submitted to
other proceedings by the authors.}:

{\it There are two dual descriptions: either one uses a short perturbative series and
adds the leading quadratic correction by hand, or one uses long perturbative series
and then there is no reason to add the quadratic correction}.

Numerically both   Eqs. (\ref{ya}) and (\ref{nora})
work well.
Chronologically, the
papers in the series \cite{linear,CNZ} appeared first. At that
time, the common belief was that the Voloshin-Leutwyler potential is valid non-perturbatively. This would
correspond to a cubic correction in Eq. (\ref{ya}) (or (\ref{nora})).
The papers \cite{linear,CNZ} established validity of the unconventional
(at that time) quadratic correction. The emphasis in later papers \cite{brambilla} was in fact on finding
another interpretation to the already known quadratic power correction.

The problem of mixing  between power-like corrections and
perturbative series is not new at all. The standard view is that  power
corrections are  related to divergences in perturbative series
due to the factorial growth of the expansion coefficients
(for review see, e.g., \cite{beneke1}).
This viewpoint formulated long time ago still dominates theoretical
thinking.  In practice, however, no factorial growth of the expansion coefficients
has been observed so far. The reason could be that the ability to calculate
the expansion coefficients is limited and the series known explicitly are not {\it long} enough.

Here we come actually to a key point. Because in phenomenological applications,
one usually assumes, explicitly or tacitly, that large-order asymptote
sets in immediately after the terms known explicitly (
 see, e.g., 
\cite{menge,JAMIN} and rererences therein).

There exists, however, an example of a long perturbative series
which allows to check the current ideas on the
expansion coefficients. We have in mind the
perturbative calculations of the gluon condensate \cite{RAKOW1,RAKOW2,BURGIO}.
This example indicates strongly that 
theorems on the asymptotic behaviour of the expansion coefficients
might apply only in case of  much longer series than are
available in reality.
Thus, we argue that, in realistic phenomenological fits, one should keep
the quadratic corrections which are absent from the
symbolic expansion in Eq. (\ref{ope}).

{\it Thus, our main point is that the
properties of the {\it relatively short} perturbative series are different
from properties of {\it long} perturbative  series.}

Another new point is the impact of the dual models.
We will argue, basing on the results of \cite{andreev1},
that   there exists another source of the quartic corrections, which are usually
identified with the infrared-sensitive part of the gluon condensate $\la G^2\ra$.
Namely, the same short-distance contributions which control the quadratic correction
taken to second order,  produce a {\it calculable} quartic correction.
We confront this insight brought by the dual models
with explicit perturbative calculations of papers in Ref. \cite{RAKOW1,RAKOW2,BURGIO}.

In Sect. 2, we discuss  an argumentation  in favor  of the  duality between the
quadratic correction  and long perturbative series.
In Sect. 3, we emphasize lessons for the generic structure of  perturbative  series
brought by the  explicit  calculations  of the  gluon condensate.
In Sect. 4, we propose a simplified generic version of  perturbative series.
In Sect. 5,
we summarize lessons
brought by   the holographic  models.
In Sect. 6, we discuss the unexpected duality between the perturbative and quartic power
corrections. Sect. 7 is devoted  to  phenomenology of particular processes.
In Sect. 8, we present  our conclusions.

\section{Duality expected (quadratic correction)}
\subsection*{\b Duality between $s$- and $t$-channels}
\nin
Because of the existing confusion in the literature concerning
the duality between long perturbative series and quadratic correction,
let us start with the notion of the duality itself.

Consider a hadronic reaction $a+b\to c+d$ at relatively low energies. Then the
following representation of the amplitude can be reasonable:
\begin{eqnarray}\label{nearest}
A(a+b\to c+d) \approx{\rm (nearest~s-channel~exchange)}~\nonumber\\
{\rm (nearest~t-channel~exchange)}~,
\end{eqnarray}
Such a phenomenology  was popular a few decades ago and turned successful.

Now, imagine that one starts improving Eq. (\ref{nearest}) by summing up the $s$-channel exchanges:
\begin{eqnarray}\label{sum}
 A(a+b\to c+d) ~\approx
\sum_{n=1}^{N~particles}\hspace*{-0.6cm} {\rm~(s-channel~exchange)}\nonumber\\ +{\rm (t-channel~exchange)}~,
\end{eqnarray}
where the sum over the s-channel resonances is taken.

Then, if $N$ is large enough one would notice that there is no more space
for the $t$-channel exchanges. The conclusion could be
that there are no $t$-channel particles or that they are decoupled
from our hadrons $a,b,c,d$.

As everybody knows, beginning with the celebrated Veneziano's
paper \cite{VENEZIA}, such a conclusion
would be wrong. Namely, if one uses sums over the resonances, then it is {\it either}
$s$-channel {\it or} $t$-channel exchanges that are allowed but not both.

 Similar things happen in the case
of the quadratic corrections to the parton model (of which a linear potential
is an example). One uses {\it either } a short perturbative series and adds a linear term
by hand. This is an analogy to the nearest-singularity amplitude in Eq. (\ref{nearest})
and corresponds to the form in Eq. (\ref{ya}). {\it Or} one uses a long perturbative series and
then {\it does not} add by hand the linear term. Since it is already included into the
perturbative series, by virtue of the general theorems inherent to the
 Yang-Mills  theories. This is then the  version in Eq. (\ref{nora}).

Thus, claiming that the parameter $\tilde{\sigma}\approx 0$
in Eq (\ref{nora}) contradicts $\sigma \neq 0$ in Eq (\ref{ya}) is
like claiming that  summing up the s-channel exchanges proves that
there are no t-channel particles in nature.

\subsection*{\b Quadratic correction and OPE}
\nin
The proof \cite{beneke} that there are no genuine non-perturbative quadratic corrections
is simple. Indeed, originally, the quadratic correction was associated with the so called
ultraviolet renormalon which corresponds to the following asymptotic series:
\begin{equation}\label{UV}
f(Q^2)_{UV~renorm.}~\sim~\sum_{N_{cr}}^\infty [\alpha_s(Q^2)]^k{(-1)^kk!( b_0)^k}~~,
\end{equation}
where $Q^2$ is a generic large mass parameter inherent to the problem and
$b_0\equiv {1\over 4\pi}(11-(2/3)n_f$ is the first coefficient  in the $\beta$-function for $n_f$ flavours.
Note that if one treats the expansion in Eq. (\ref{UV}) as an asymptotic series,
then its uncertainty is  a quadratic  correction,
$\Lambda_{QCD}^2/Q^2$.
On the other  hand one can sum up  the series \`a la Borel
and then there is no  uncertainty at all.

The crucial observation is that the factorial growth of the expansion coefficients
in Eq. (\ref{UV}) are associated with an integration over very large momenta, $p^2\gg Q^2$.
However, because of the asymptotic freedom, this region of integration should  not
be a source of uncertainty in QCD. Indeed, by introducing a cut off $a$
and using the coupling $\alpha_s(a)$ normalized in the UV,
one eliminates the integration over momenta $p^2>a^{-2}$ and, therefore,
there is no ambiguity of order $1/Q^2$ \cite{beneke}.

From the point of view of the operator product expansion (OPE) in
Eq. (\ref{ope}),  the quadratic correction we are discussing is hidden
in the coefficient function in front of  the unit operator, $C_I(\alpha_s(x))$
and, in no way, violates the OPE.

However,
the QCD (spectral) sum rules were originally based \cite{svz}
(for a review, see e.g. \cite{SNB}) on a simplified
assumption that the coefficient functions can be approximated by the first terms,
while the effect of the confinement is encoded in the power corrections.
It is only
{\it within this terminology} that one might say that the form in Eq. (\ref{ya}) violates
the OPE. In a more correct but longer language, what is violated is
the assumption  that the coefficient
functions
are approximated by their first terms. More advanced applications
of the sum rules are keeping longer and longer perturbative series.
Then, the terminology with `violations of the OPE' due to the quadratic correction
becomes obsolete.

Another source of confusion is the observation that,
in the Euclidean space, one can ascribe a gauge-invariant meaning
to the vacuum matrix element of dimension-two operator $(A_{\mu}^a)^2$ \cite{kondo}.
The quantity $\langle 0|(A_{\mu}^a)^2|0 \rangle_{min}$
turns to be of significant interest in many applications. This  does  not change of
course the fact
that $\langle 0|(A_{\mu}^a)^2|0 \rangle_{min}$ does not appear in the operator product expansion (\ref{ope}).
None of the papers in Ref \cite{linear} claims either violation of the OPE with
`long' perturbative expansions or appearance of $\langle 0|(A_{\mu}^a)^2|0 \rangle_{min}$
in the OPE equations. Nevertheless, sometimes one fights just with these
would-be-made claims \cite{brambilla}.

From the perspective of the perturbative expansion, the most difficult question is: why
the quadratic correction could be at all important ?
Therefore, it is worth emphasizing that the quadratic correction is required
by phenomenology, not yet by the theory.
For example, on the lattice, one can give a definition of  the `non-perturbative'
heavy quark potential (for review see e.g. \cite{zakharov}).
Then, this potential is pure linear starting from the smallest
distances available:
\begin{equation}
V_{Q\bar{Q}}(r)\vert_{non-pert}~\approx~\sigma\cdot r~.
\end{equation}
Moreover, this non-perturbative contribution encodes confinement as well.
Thus, it is strongly suggested by the phenomenology that the
effects of the confinement are encoded in the quadratic correction
which is not explicit in the general OPE in Eq. (\ref{ope})
\,$^2$\footnotetext[2]{A phenomenologically successful fit to the power corrections is
provided by the `short-distance' gluon mass (see \cite{CNZ,SNe,SNtau,SNmass,ARRIOLA}).
However, the very notion of the short-distance gluon mass can be introduced only
in the Born approximation and only for a certain class of processes \cite{linear,CNZ}.}.
As we argue in Section {\ref{sec:holography}},  a natural framework for
the quadratic correction  is provided by the stringy, or holographic formulation
of QCD
(for a review see, e.g.,  \cite{klebanov}) .
\section{Lessons from PT calculation of $\langle {\alpha_s}GG\rangle$}
\nin
The best check of this logic is provided by the beautiful results for perturbative
calculation of the gluon condensate on the lattice (the most advanced calculations are due to Rakow et al. \cite{RAKOW1,RAKOW2,BURGIO}):
More precisely, the results refer to perturbative evaluations of the quantity:
\beq\label{condensate}
a^4{\pi\over 12N_c}\Big[{-b_0g^3\over\beta(g)}\Big]\langle {\alpha_s}GG\rangle
= 1+\sum_{n=1}^Np_ng^{2n}+\Delta_N~,
\eeq
where $a$ is the lattice spacing and $\alpha_s(a)$ is the running coupling normalized at the
ultraviolet cut off, $p_n$ are the  expansion coefficients which are calculated explicitly
up to $n=N$. Finally the difference $\Delta_N$ is known numerically since the total value
of the l.h.s. of Eq. (\ref{condensate}) co\"\i ncides with the plaquette action and is known to a very
good precision. Moreover, the difference is fit to power-like corrections:
\begin{equation}
\Delta_N~=~b^N_2(\Lambda_{QCD}\cdot a)^2+b^N_4(\Lambda_{QCD}\cdot a)^4~~,
\end{equation}
where the coefficients $b_{2,4}$ are fitting parameters
which depend on the number of perturbative terms calculated explicitly.

Explicit results \cite{RAKOW1,RAKOW2,BURGIO} demonstrate that, indeed, the power corrections
in Eq (\ref{condensate}) depend strongly on the number $N$ of perturbative terms
taken into account explicitly. Namely, up to $N\approx 10$ the power corrections are
dominated by a quadratic term:
\begin{eqnarray}\nonumber
\Delta_N~\approx~b_2^N \cdot(a\cdot \Lambda_{QCD})^2~~~~{\rm for}~~~~N\leq 10~.
\end{eqnarray}
That is, the coefficient $b_4^N$ is consistent with zero for such $N$.
However, the numerical value of the coefficient $b_2^N$ in front of the power correction
diminishes with increasing $N$. Thus, perturbative corrections `eat up'  the
power correction. In more refined terminology, the perturbative terms
are dual to the leading power correction.

At $N>10$, a quartic correction emerges as a result of subtracting the
perturbative contributions from the total matrix element $\langle \alpha_sG^2\rangle$:
\beq
\label{quartic}\nonumber
\Delta_N~\approx~const \cdot(a\cdot \Lambda_{QCD})^4 ~~~~{\rm for}~~~~N\geq 10~.
\eeq
And, finally, at about $N\approx 16$ one restores the value of the quartic
correction which is  \cite{RAKOW1,RAKOW2}:
\begin{equation}
\langle{\alpha_s}G^2 \rangle_{pert}~\approx~0.12~{\rm GeV}^4,
\end{equation}
with a large error, but the result is comparable in magnitude with the standard gluon condensate entering
the QCD (spectral) sum rules.
Another remarkable finding \cite{RAKOW1,RAKOW2} is that perturbative coefficients $p_n$
entering (\ref{condensate}) are well approximated by a simple geometric series:
\begin{equation}
r_n~\equiv~{p_n\over p_{n+1}}~=~u\Big(1-{1+q\over n+s}\Big)~,
\label{ratiocoeff}
\end{equation}
where the fitting parameters $u=0.961(9),~q=0.99(7),~s=0.44(10)$.
The perturbative series with such coefficients is convergent for
$$|g^2|~<~|u|^{-1}~~.$$
This simple geometrical series fits explicit calculations of the PT coefficients at least for the
first 16  terms. Extending $n\to \infty (n\geq 50)$, the geometric series  reproduces the
full answer to the accuracy  better than $10^{-3}$, which is
a remarkable result.

\section{Geometric growth of the PT coefficients ? }
\label{sec:geometric}
\nin
Physicswise, one can say that the series found in \cite{RAKOW1,RAKOW2,BURGIO}
is determined by  the singularity due to the crossover from strong to  weak coupling.
This  is true in pure gluonic  channel. This could also be true with account
of quarks. Then, we would  have, in different channels, geometric series,
with approximately  the  same range of convergence.
To  see whether  a such hypothesis can be ruled out,  we compile
below the calculated expressions of the Adler-like function in the euclidian region\,\footnote{One can notice that the PT corrections in the theory of $\tau$-decay:
$
\delta^{(0)}=a_s+5.202a_s^2+26.366a_s^3+127.079a_s^4
$
\cite{BNP,LEDIB,PIVO,KUHN} indicates a geometric growth, but the effects due to the analytic continuation and to the
the $\beta$-fuctions induced by the renormalization group equation obscure the exact behaviour of the coefficients. Interpretations of a this fact require more involved analysis (see e.g. \cite{KATAEV}).}
for
different channels.

In the vector channel with massless quarks, it reads \cite{DINE,LARIN,KUHN}:
\bea
-Q^2 {d\over dQ^2}\Pi_V(Q^2)&=& {N_c\over 12\pi^2}\big{[} 1+a_s+1.64a_s^2+\nnb\\
&&6.31a_s^3+49.25a_s^4\big{]}~.
\eea
The perturbative corrections to this expression due to the strange quark mass for the neutral (resp. charged) vector
current, read \cite{BNP,CHETms}:
\bea
Q^2\Pi_{\bar ss}^{D=2}= -{6\bar m_s^2\over 4\pi^2}\Big{[} 1+2.67a_s+24.14a_s^2+250a_s^3\Big{]}~,\nnb\\
Q^2\Pi_{\bar us}^{D=2}= -{3\bar m_s^2\over 4\pi^2}\Big{[} 1+2.33a_s+19.58a_s^2+202a_s^3\Big{]},
\eea
The difference from $\alpha_s^2$ is due to the light by light
scattering diagram contributing in the neutral vector two-point correlator.

For the pseudoscalar channel, the QCD expression of the Adler-like function reads for $n_f=3$ \cite{BECCHI,LARIN1,CHET2}:
\bea
-Q^2 {d\over dQ^2}\Pi_5(Q^2)&=&{N_c\over 8\pi^2}\Big{[}1+5.67a_s+45.85a_s^2+\nnb\\
&&465.8a_s^3+5589a_s^4\Big{]}~.
\eea
Similarly, one can also present the PT expressions of moments in deep-inelastic scatterings. The ones of  the well-known Ellis-Jaffe \footnote{The Bjorken sum rule corresponds to: $\int_0^1dx~ (g_1^{p}-g_1^{n})$.} for polarized  electroproduction or of the Gross-Llewellyn Smith sum rule for neutrino-nucleon scattering, read, for $n_f=3$ \cite{KODAIRA,LARIN2}:
\bea
\int_0^1dx~ g_1^{p(n)}\simeq (1-a_s-3.58a_s^2-20.22a_s^3)\Big{[}\pm {|g_A|\over 12}\nnb\\ 
+{a_8\over 36}\Big{]}
-{a_0\over 9}(1-0.33a_s-0.55a_s^2-4.45a_s^3)~,\nnb\\
\int_0^1 dx~F_3^{\bar \nu p+\nu p}\simeq 6(1-a_s-3.58a_s^2-18.976a_s^3)~,
\eea
where $a_8$ and $a_0$ are the octet and singlet structure functions.
One can notice that, in all the cases, the series found, do not show any factorial growth nor an alternate sign but, are consistent with geometric
series, with sizable corrections at small $n$ similar to  the case of the gluon condensate\,\footnote{A geometric growth of the PT coefficient has been assumed in \cite{LEDIB} for predicting the $\alpha_s^4$ term of the PT series of the D-function in the V+A channel. This result has been (approximately) confirmed later on by the analytic calculation of \cite{KUHN}.} .
Thus, there  is an exciting  perspective that  all  the perturbative series are in fact
quite simple in large orders.
\section{Insight from dual models} \label{sec:holography}
\subsection*{\b Holographic quadratic correction}
\nin
In the holographic language, one evaluates the same observables, as within the
field theoretic formulation of QCD, but in terms of strings living in extra dimensions.
There is no direct derivation of the metrics of the extra dimensions in
the QCD case. One rather uses phenomenologically motivated assumptions
(see, e.g., \cite{andreev1}).

The crucial element is the metrics $z$ in the fifth dimension.
Following \cite{andreev1}, let us choose the following model:
\begin{equation} \label{simple}
ds^2=R^2{h(z^2)\over z^2}(dx_i^2+dz^2),
\end{equation}
where $R^2$ is a constant whose explicit definition is not important
for us here and the function $h(z^2)$ is specified below. Note that, at $z\to 0$, one needs
$h(z^2)\to  const$ in order to reproduce an approximate conformal symmetry of the Yang-Mills
theories (due  to the asymptotic freedom).

We would like to define the function $h(z^2)$ in such a way
as to ensure confinement at large distances and to reproduce the (leading) quadratic
power correction at short distances. The following choice:
\begin{equation}\label{assumption}
h(z^2)=\exp(c^2z^2/2)~~
\end{equation}
satisfies these conditions.
Note that, while the condition to reproduce confinement, or the area law for the
Wilson line is common to all the holographic models, the condition to reproduce
the quadratic correction at short distances assumes that it is  this correction
which  encodes  the  confinement at short distances. One can demonstrate
that, assuming  Eq. (\ref{assumption}), is equivalent to  assuming the Cornell potential
for the heavy quarks interaction. The numerical  value of  the constant $c$
can be fixed in terms of the string tension, $c^2=(0.9~{\rm GeV})^2$.

The simple model in Eqs (\ref{simple}) and (\ref{assumption})
turns to be successful phenomenologically (see, in particular, \cite{andreev2}
and references therein).

A crucial advantage of using the hologhraphic language is that
it allows for a perfectly gauge invariant  way to introduce and parameterize
the quadratic correction. Also, the simple expression (\ref{assumption})
looks much more `natural' than the assumption on approximate equality of the long
perturbative series and quadratic correction plus short series 
discussed above.
What is lacking, is further applications of the same metrics in Eq. (\ref{assumption})
to evaluate quadratic corrections to the parton model in other cases, such as the current correlators.

\subsection*{\b Holographic quartic correction}
\nin
Presence of the quadratic correction in the string-based approach
is an assumption which allows to  model the metric in the fifth
dimension. However, once the metric is fixed, one can calculate
the full answer for the gluon condensate \cite{andreev3}.

The model does  not account for the running of the coupling but allows to
evaluate power corrections. In particular, it produces the value of
the `physical gluon condensate' of the magnitude:
\begin{equation}
\langle{\alpha_s} G^2\rangle_{holographic}~\approx~0.03~GeV^4,
\end{equation}
which is  reasonable phenomenologically \cite{svz}.

What appears even more important is that the dual-model approach
provides a new qualitative picture for the power corrections.
Namely, in the holographic language $\la\alpha_sG^2\ra\sim\Lambda_{QCD}^4$
appears as a second-order effect in the coefficient $c$ introduced in Eq. (\ref{assumption}):
\begin{equation}\label{short}
\la\alpha_sG^2\ra_{holographic}~\sim~c^2~\sim~\Lambda_{QCD}^4~.
\end{equation}
Since the coefficient $c$ (or the quadratic correction in the holographic language)
is associated with short distance, the same is true for the
gluon-condensate contribution in Eq. (\ref{short}).

In short, the stringy calculation does not have a counterpart to
the infrared-renormalon contribution which is taken for granted
in field theoretic approach. This point is worthy to be elaborated.

In both cases of field theory and of stringy calculation,
one deals with a propagator, of a particle or a string respectively.
In both cases, the leading contribution comes from short distances.
If the typical size is of order $a$, then, in both cases, $\la \alpha_s G^2\ra\sim a^{-4}$.
However, the probability for a (virtual) particle
to propagate to the distance of order $\Lambda^{-1}_{QCD}$
is power-like  suppressed:
\begin{equation}\label{particle}
a^{4}\la\alpha_sG^2\ra_{IR,particle}~\sim~(\Lambda_{QCD}\cdot a)^4,
\end{equation}
as revealed by the infrared renormalon (see, e.g., \cite{beneke1}).
In the case of strings, the suppression of the infrared region turns to be
exponential:
\begin{equation}
a^{4}\la\alpha_sG^2\ra_{IR,string}~\sim~\exp\ga -const/\big{[}\Lambda_{QCD}\cdot a\big{]}^{\gamma}\dr,
\end{equation}
where $\gamma $ is positive.
Intuitively, this strong suppression  is due to the
fact that string corresponds to a collection of particles.

\section{Duality unexpected: quartic correction}
\nin
  Let us emphasize again that the standard assumption
  is that  the quartic correction
in Eq. (\ref{quartic}) emerges simultaneously with the factorial divergence
in expansion coefficients $a_n$ (see
Eq. (\ref{condensate})):
\begin{equation}\label{divergence}
\Big({p_{n+1}\over p_n}\Big)_{IR~renormalon}~\sim ~n~~~{\rm for}~~~n\gg~1~.
\end{equation}
This divergence is due to the infrared renormalon
(for a review, see \cite{beneke1}).

 So far \cite{RAKOW1,RAKOW2,BURGIO}, one does not run into the
problem of the divergence in Eq. (\ref{divergence}):
\begin{equation}
\Big({p_{n+1}\over p_n}\Big)_{n<15}~\sim~1~.
\end{equation}
It is even more amusing that, with presently available
perturbative terms in Eq (\ref{condensate}),  one can extract
\cite{RAKOW1,RAKOW2,BURGIO} the
`genuine' gluon condensate in Eq. (\ref{quartic}):
\begin{equation}\label{rakow}
a^4\la\alpha_sG^2\ra~\sim~(\Lambda_{QCD}\cdot a)^4~~,
\end{equation}
so that the quartic correction gets disentangled from the
infrared renormalon. This observation, if confirmed, is a radical change of dogma.

It is not ruled out that the infrared renormalon still shows up
in higher orders of perturbation theory, say, at $n\sim 25$,
as discussed in \cite{RAKOW1,RAKOW2}. However, its contribution will be in any case
smaller than the condensate in Eq. (\ref{rakow}) determined from perturbative series
which looks like a geometric series and exhibits no factorial
growth of the coefficients.

It is amusing that the dual models independently provide
a mechanism of generating the quartic correction from short
distances {[}see discussion in Eq. (\ref{condensate}){]}. The condensate
in Eq. (\ref{condensate})  is not related to any divergence of the
perturbative theory either. Thus, two  independent approaches result
in similar  pictures.

\section{Phenomenology of $1/Q^2$ corrections}
In this section we review results of numerical fits which
keep both a few-term perturbative series and a qudratic correction
(assuming, therefore that considerably more perturbative terms
are needed to apply the duality).
\nin
\subsection*{\b Tachyonic gluon mass squared $\lambda^2$}
\nin
From  the  phenomenological point of view, it would  be important
to  relate the quadratic  corrections  in various channels.
A model  which turns successful in this respect is the introduction of
a tachyonic
gluon mass $\lambda^2$ at short distances \cite{linear,CNZ}. From the  calculational
point of view one changes  the gluon propagator:
\beq\label{propagator}
D_{\mu\nu}^{ab}(k^2)={\delta^{ab}\delta_{\mu\nu}\over k^2}\rar {\delta^{ab}\delta_{\mu\nu}\over k^2}\ga 1+{\lambda^2\over k^2}\dr~.
\eeq
and checks that the quadratic correction is associated with large momenta $k^2\sim Q^2$.
To the lowest order the analysis is gauge invariant.  The model in Eq. (\ref{propagator}) is purely heuristic in nature
and can be used only for estimates in conjunction with short
perturbative series.
\subsection*{\b Estimate of $\la \alpha_s G^2\ra$ and  of $\alpha_s\lambda^2$ }
\nin
One can extract these two parameters by using ratio of exponential QCD (spectral)
sum rules in $e^+e^-\to$ hadrons data \cite{SNe}, which is not sensitive to
the leading $\alpha_s$ corrections. It is worth mentioning that FESR may not be
appropriate for extracting such small quantities, as it requires a cancellation of
two large numbers which depend on the high-energy
parametrization of the spectral function. This feature is signaled by the large
range spanned by the determinations of power corrections
using FESR \cite{PEROTTET} and  the discrepancies   of the estimated quadratic corrections in \cite{domi1}
and \cite{domi2}, which both also differ from the one using the ratio of exponential Borel/Laplace (LSR) used in \cite{SNe}.

In addition to previous channels, the gluon condensate can be also obtained using a ratio of LSR for the $J/\psi-\eta_c$ and $\Upsilon-\eta_b$
mass-splittings, which has a minimum sensitivity on the heavy quark mass effects and on the $\alpha_s$
corrections \cite{SNh}
\footnote{ Some estimates of $\la \alpha_s G^2\ra$ in the existing  literature suffer
from  correlation with $\alpha_s$ and $m_Q$.
We plan to reanalyze these sum rules.}.

The resulting values of the parameters are \cite{SNe,SNtau,SNh,SNB}:
\bea
 \la \alpha_s G^2\ra&=& (6.8\pm 1.3)10^{-2}~{\rm GeV}^4~,\nnb\\
a_s\lambda^2&=&-(6.5\pm 0.5)10^{-2}~{\rm GeV}^2~,
 \eea
 where $a_s\equiv  {\alpha_s/\pi}$ and where the value of the gluon condensate is about
2 times the original SVZ value as expected from Bell-Bertlmann analysis
\cite{BELL} \footnote{A detailed comparison with the SVZ result can be found in \cite{SNB}.}.

 One can also use the  pseudoscalar LSR for extracting $a_s\lambda^2$ \cite{CNZ}.
Studying the stability of $(m_u+m_d)$ with respect to the change of $\lambda^2$, one obtains, at the stability region in $\lambda$,
 a reduction of the value of light quark mass of about 5\% and the corresponding $\lambda$-value:
 \beq\label{eq:lambda2}
 a_s\lambda^2=-(12\pm 6)10^{-2}~{\rm GeV}^2~,
 \eeq
 consistent with previous estimate from $e^+e^\to$ hadrons data though less accurate.
Taking into account these uncertainties,  we shall consider the conservative value:
 \beq\label{eq:lambda}
 a_s\lambda^2=-(7\pm 3)10^{-2}~{\rm GeV}^2~.
\eeq
 These results indicate that these power corrections are small though crucial for
understanding the non-perturbative properties of QCD. One can also notice that the new
quadratic correction can only slightly change the existing QSSR phenomenology because of its smallness.
 \subsection*{\b QCD (spectral) sum rule scales}
 \nin
Remarkably enough, the simple model in Eq. (\ref{propagator})
brings in a qualitative success, explaining various
mass scales (details can be found in the original paper \cite{CNZ})
revealed by analysis of the sum rules \cite{NOVIKOV}.
To our knowledge, there is no alternative explanation of the numerical hierarchy of  such different scales

 \subsection*{\b Light quark masses}
\nin

 \vspace*{-0.5cm}
{\scriptsize
\begin{table}[hbt]
\setlength{\tabcolsep}{0.4pc}
 \caption{\scriptsize    QSSR predictions of the light quark masses in units of
MeV to order $\alpha_s^3$ and including the $1/Q^2$ correction. For translating $(m_u+m_d)$  into $m_s$ (and vice-versa), we have used the ChPT prediction for  $m_s/(m_u+m_d)$ . }
 {\small
\begin{tabular}{lllll}
\hline
Channels&$(\bar m_u+\bar m_d)(2)$&$\bar m_s(2)$&Ref.   \\
\hline
LSR Pion&8.6(2.1)&$\lrar$&$ 107.4(22.0)$&\cite{CNZ,SNB}\\
LSR Kaon&--&&119.6(18.4)&\cite{SNB}\\
$\tau$-decay&--&&$93(30)$&\cite{SNms}\\
$e^+e^-$&--&&$104.3(15.4)$&\cite{SNmass}\\
%
{\bf Average}&{\bf 8.7(1.3)}&{\boldmath$\llar$}&{\bf 106.2(15.4)} \\
\hline
\end{tabular}
}
\label{tab:mass}
\end{table}
}
 \vspace*{-0.5cm}
\nin
Effects of quadratic corrections in the determinations of the light quark masses
have been studied in \cite{CNZ,SNB,SNmass}. As anticipated previously,
the absolute value of $|\lambda^2|$ tends to lower by 5-6\% the values of $(m_u+m_d)$ and of
$m_s$ running masses obtained from (pseudo)scalar sum rules, while it tends to increase
the value of $m_s$ in $e^+e^-$ and in $\tau$-decay data. These different results and their first published average are summarized in Table \ref{tab:mass}.
%

These results are in better agreement with some recent lattice calculations based on non-perturbative normalization than the recent global average given in \cite{SNB,SNmass} which includes determinations without the $1/Q^2$-term.

 \subsection*{\b $\alpha_s$ from $\tau$-decay}
 \nin
One of the sensitive places where the effect of the quadratic term can be important is the precise
extraction of $\alpha_s$ from $\tau$-decays \cite{BNP,LEDIB}.
One of the authors has presented recently \cite{SNtau}
 analysis of the effect of the quadratic corrrection on the
 determination of $\alpha_s$. The result is:
\bea
\als(M_\tau)&=& 0.3249~(29)_{\rm ex}(9)_{\rm st}(74)_{\rm nst} \lrar\nnb\\
\als(M_Z)\vert_\tau&=&0.1192~(4)_{\rm ex}(1)_{\rm st}(9)_{\rm nst}(2)_{\rm ev}~,
\label{eq:alfaaverage}
\eea
where the errors are due respectively to the data, to the standard and non-standard corrections and to
the evolution from $M_\tau$ to $M_Z$.
This value of $\alpha_s$ is in agreement with existing estimates \cite{menge,KUHN,DAVIER,ALPHAS,PERIS} obtained using different
appreciations of the non-perturbative contributions and of the large order perturbative series. This result agrees with the ones from the $Z$-width \cite{KUHN} and from a global fit  of electroweak data at ${\cal O}(\alpha_s^4)$~\cite{DAVIER}:
\beq
 \alpha_s(M_Z)\vert_{N^3LO} =  0.1191~ (27)_{\rm exp}(1)_{\rm th}~,
\label{eq:alphamz}
\eeq
and with the most recent world average \cite{BETHKE}:
\beq
\als(M_Z)\vert_{\rm world}=0.1189~(10)~.
\label{eq:alphaworld}
\eeq
One can notice that the $1/Q^2$ contribution tends to decrease the value of $\alpha_s$ obtained without this term and improves the agreement with the world average.
\section{Conclusions}
\nin
In conclusion, our main point here is that large-order
perturbative and non-perturbative contributions mix  up
as a matter of principle. The duality between these corrections
is  expected theoretically.

The duality, however, was thought  to  be confined to the quadratic  corrections.
The most recent and intriguing development is that this perturbative - nonperturbative
duality might extend to the quartic correction as well. Basing on the existence
of the infrared renormalon in perturbation theory, one would not expect that
the quartic correction is calculable via the long perturbative series.
Therefore, it is a challenge
to explain the numerical observations on the perturbative series.
\cite{RAKOW1,RAKOW2,BURGIO}.

The holographic approach \cite{andreev1} does suggest a mechanism
for  generating quartic corrections at short distances
but much more is needed to be done to finally clarify the issue.
In the holographic language the quadratic correction looks as a stringy correction
\footnote{Further support for this result comes from lattice studies of confinement (for a review, see, e.g., \cite{zakharov}).}.

Taken at face values, these observations accumulate to a drastic change of expectations
on behaviour of perturbative series at higher orders in pure gluonic sector.
Instead of factorial divergences in the expansion coefficients and related power-like
terms, there are emerging convergent and calculable series , or dual
to the power-like terms. 
We scrutinized the newly emerging picture against the phanomenology
and did not find any flaws.

Existence of the infrared renormalon has never been proven since
it corresponds only to a subclass of all the graphs in a given order
of the perturbation theory. However, within the field-theoretic formulation
it is equally difficult to imagine that these graphs are cancelled.
The dual, or stringy formulation provides an alternative view. Within
these models it is the power corrections which are calculable most directly.
They are coming from short distances. Therefore they should be calculable
perturbatively within the field theory and one gets explanation why
the infrared-renormalon graphs cancel.

Thus, small steps in the phenomenological analysis of the power corrections
might accumulate to produce a new insight to the fundamental issues 
of QCD. It goes without saying that further checks of the novel picture
are needed. 

\section*{Acknowledgements}
\nin
The authors are grateful to E. Ruiz Arriola, M. Golterman, S. Peris and A. Pineda for discussions and to K. Chetyrkin and P. Rakow for communications. We also thank Santi Peris for the invitation at the UAB (Barcelona). V.Z.  thanks the LPTA (Montpellier) for the hospitality. S.N. thanks the partial support of the CNRS-IN2P3 within the QCD and Hadron Physics program.

\end{document}